# High-resolution surface structure determination from bulk X-ray diffraction data

Nirman Chakraborty and Swastik Mondal[*]

*CSIR-Central Glass and Ceramic Research Institute, 196, Raja S. C. Mullick Road, Jadavpur, Kolkata-700032, India*

[*]Corresponding author: swastik_mondal@cgcri.res.in

**Abstract:**

The key to most surface phenomenon lies with the surface structure. Particularly it is the charge density distribution over surface that primarily controls overall interaction of the material with external environment. It is generally accepted that surface structure cannot be deciphered from conventional bulk X-ray diffraction data. Thus, when we intend to delineate the surface structure in particular, we are technically compelled to resort to surface sensitive techniques like High Energy Surface X-ray Diffraction (HESXD), Low Energy Electron Diffraction (LEED), Scanning Transmission Electron Microscopy (STEM) and Grazing Incidence Small Angle X-ray Scattering (GISAXS). In this work, using aspherical charge density models of crystal structures in different molecular and extended solids, we show a convenient and complementary way of determining high-resolution experimental surface charge density distribution from conventional bulk X-ray diffraction (XRD) data. The usefulness of our method has been validated on surface functionality of boron carbide. While certain surfaces in boron carbide show presence of substantial electron deficient centers, it is absent in others. Henceforth, a plausible correlation between the calculated surface structures and corresponding functional property has been identified.

Several physico-chemical processes proceed via interaction of external factors with the material's surface [1-3]. For example, in gas sensing, it is the interaction of target gas molecule with material's surface which determines the result of a sensing phenomenon [4-5]. For catalysis and dye degradation, it is the mutual interaction of target molecule with the catalyst surface which determines the route in which the reaction may proceed [6-7]. Similar procedures are followed in other processes like $H_2$ generation by photo catalytic water splitting, solar energy harvesting, gas separation through membranes and water purification by removal of heavy elements [8-10]. The wide applicability of surface mediated phenomena makes the role of material's surface very important and decisive, preparing scope for detailed analysis of surface and surface structures in functional materials [1-5]. Understanding surface structure has remained a major concern for scientists since decades. Consequently, there have been several definitions of surfaces of solids. Surface has been defined as a particularly simple type of interface at which the solid is in contact with the surrounding world, i.e. the atmosphere or in ideal case, the vacuum [11]. In crystallographic terms, surface is an entity which differs from the bulk by the fact that the periodic unit cell stacking is finite in one of the three dimensions [12]. This allows the near-surface unit cells to modify their geometric arrangements to give rise to a specific surface structure [12]. However, while the surface is an inevitable entity in every material system, till date the tools to elucidate surface charge density are very limited.

While X-ray diffraction remains as the most popular and effective tool for identifying and analyzing bulk crystal structures [13-17], the processes involved for determining surface structures have been perceived to be complex and has been both considered and executed separately from bulk XRD studies [18]. Gustafon *et al*. devised a method of analyzing surfaces by high-energy surface X-ray diffraction (HESXD) using photons of energy 85 keV [19]. Several other works on surface structure analysis have been conducted by the methods of Low Energy Electron Diffraction (LEED) and diffused LEED (DLEED) [20-21]. However, technical problems with all the methods mentioned above except X-ray diffraction involve special and elaborate experimental facilities with considerable expertise in data interpretation as well [20-21]. While the experiments are too sophisticated and unavailable to most researchers, the complication with their execution adds significantly to above difficulties. A recent development in this respect has been introduction of real space imaging technique to map local charge density of crystalline materials in four dimensions with sub-angstrom resolution for imperfect structures like interfaces, boundaries and defects [22]. However, the drawbacks of the technique lie in complexity and availability of experimental set-up which involves scanning transmission electron microscopy (STEM) alongside an angle-resolved pixellated fast-electron detector [22]. Grazing Incidence Small Angle X-ray Scattering (GISAXS) is another technique to estimate surface structures [23-24]. However, the utility of this method has been till date confined to thin films and nanostructures only. This calls for need to develop a convenient complementary method by which elucidation of surface structure (charge density distribution) with higher precision is feasible.

The charge density at any point is determined by the contribution of charge densities of surrounding atoms/sources [25-26]. Using the concept of Green's function, the contributions from different atomic basins towards a particular point can be calculated based on the formula [25-26]

$$\rho(\boldsymbol{r}) = \int \frac{-\nabla^2 \rho(\boldsymbol{r}')}{4\pi|\boldsymbol{r}-\boldsymbol{r}'|} dr' \ldots\ldots\ldots\ldots (1)$$

Where $\rho(\boldsymbol{r})$ is the charge density at position $\boldsymbol{r}$ due to densities at other points indicated by $\rho(\boldsymbol{r}')$. As evident from equation (1), if the contributions of atoms above a surface towards a point on the surface are not included, then the charge density calculated at that point should be different than that in bulk. This indicates that when surface truncation occurs, due to above explanation, the charge density distribution in the truncated surface will be different than that of bulk [18, 27]. Thus if contribution of charge density sources above a plane (surface) could be excluded and rearrangement of charge density due to that exclusion could be made, then in-principle, surface charge density can be obtained from bulk charge density.

In an X-ray diffraction experiment from a crystal, the diffracted intensity is the time averaged scattering intensity of all electrons in the crystal [28]. In principle one can thus obtain time averaged charge density distributions in bulk and in turn on a plane. If this plane is considered to be the surface, then technically, after removing the charge density contributions from the atoms above the plane, surface structure can be determined from bulk X-ray diffraction. However, this was never realized, as in conventional Independent Atom Model (IAM), the atoms are considered as discrete and independent of one another; neglecting deformation of charge densities due to atom-atom interactions [29]. As a result there cannot be any difference between the charge density distribution in bulk and surface in IAM. During the 2$^{nd}$ half of last century, the development of methods for aspherical charge density analysis changed this picture where aspherical charge density models such as multipole (MP) formalisms have been introduced [30-32]. One such formalism by Hansen and Coppens has been expressed as [30]:

$$\rho_{at}(\mathbf{r}) = P_c \rho_{core}(r) + P_v k^3 \rho_{valence}(kr) + \sum_{l=0}^{l_{max}} k'^3 R_l(k'r) \sum_{m=0}^{l} P_{lm\pm} d_{lm\pm}(\theta, \varphi) \ldots\ldots(2)$$

where the density of each atom ($\rho_{at}$) is divided into core, valence and deformation parts. The parameters $P_c$ and $P_v$ in first two terms of the equation denote the core shell and valence shell population coefficients [30]. The third term stands for the aspherical features of valence charge density, introducing the idea of deformation of charge density due to chemical bonding and other inter-atomic interactions [30]. The concept of bonding between atoms via charge transfer, otherwise ignored by IAM was thus included in such aspherical charge density formalisms [30]. Hence, using aspherical charge density models, the charge density contributions of an atom on other atoms could be calculated and excluded, making the formalism relevant to determining surface of a material [30]. The present work devises a simple method by which experimental surface charge distribution of materials can be determined from conventional bulk XRD data, employing concepts of aspherical charge density analysis [30-32]. An idea has been introduced as how to utilize information obtained from above studies in correlating surface structures with associated surface properties.

Charge density models based on low temperature, high-resolution single-crystal X-ray diffraction data of molecular solid α-glycine and extended solids α-boron, boron carbide and calcium mono-silicide (CaSi) were taken as the test cases for this study (see Supplementary Information for crystal structures) [33-37]. Accurate multipole (MP) models have been derived for all samples using high-resolution single crystal X-ray diffraction data from bulk crystals [33-37]. For calculating the surface charge density of a particular solid, three atoms in the solid were chosen in such a way that they constitute a triangle (Fig. 1). Then the geometrical centre of the triangle was identified. Depending on the values of unit cell lengths and keeping the centre of the analyses at the above geometrical centre, an $x \times y \times z$ (x: length extending along x axis, y: length extending along y axis and z: length extending along z axis) block of the crystal was considered.

Then with respect to the plane defined as midway of a particular crystallographic axis, the whole block was considered to be composed of two segments: one segment consisting of the plane and the region below it; other is the whole block with the plane midway (Fig. 1).Topological properties [38] like critical points on a definite plane were calculated using the XDPROP module of XD2006 software [39]; first with the segment consisting of the plane (surface) and the region below it, and secondly for the whole block (bulk) (Fig. 2). Then, the difference in values of charge density (ρ) in $e/Å^3$ between both the above cases were calculated (Table 1). The above calculations were conducted both considering atom-atom interactions using the multipole (MP) formalism and avoiding atom-atom interaction using IAM [37]. Final charge density plots were generated using XDGRAPH module of XD2006 software [39] to visualize the surface from a perspective of difference in charge density plots (Fig. 3-4 and Supplementary Information).

Charge density studies for boron carbide [36, 40] conducted on a plane drawn by central B(3) atom of a C(4)-B(3)-C(4) chain and two carbon atoms from a neighbouring chain *(C(4)-X1_B(3)-X5_C(4))* forming a triangular pattern reveal difference in charge density values between the two segments, therefore elucidating charge density distribution patterns in a surface containing C(4), X1_B(3) and X5_C(4) atoms (Fig. 3). For the C(4)-B(3) bond, a difference in charge density value of 0.001 $e/Å^3$ was found at the bond critical point (Table 1) [38]. Whereas for bond between X5_C(4) and B(3) atoms; the difference is 0.033 $e/Å^3$ at the corresponding bond critical point (Table 1). For other bonds on the studied plane like those between X1_C(4), X1_B(3); X1_C(4), B(3), the differences in charge density values were calculated to be 0.002 $e/Å^3$ (Table 1) [22]. And, for the plane drawn using two polar and one equatorial boron atoms *(X4_B(2)-X5_B(2)-X6_B(2))* [36], the difference in charge density values at critical points has been calculated to be negligible (Table 1 and Supplementary Information). Figure 3(a-d) demonstrates the charge density and deformation density maps of plane drawn using C(4), X1_B(3) and X5_C(4) atoms in the boron carbide system with difference in charge density marked as red contour lines at the surface.

Analysis on a plane drawn using *(X3_B(1), B(1), X2_B(1))* atoms using multipole (MP) formalism for intra-icosahedral interactions of α-boron [35] reveals that between the atoms X2_B(1), X3_B(1) and B(1), there exists a ring critical point [38] (Table 1). Difference calculation between charge density values of two segments has been found to be 0.014 $e/Å^3$. In between atoms X3_B(1), B(1); X2_B(1), B(1) and X2_B(1), X3_B(1) there exists a bond critical point [38] with difference charge density of 0.016 $e/Å^3$. However for the bond between two asymmetrical boron atoms B(1) and B(2), the difference is 0.004 $e/Å^3$ (Table 1). For bond between X5_B(2), B(2); the difference is 0.005 $e/Å^3$. For atoms X5_B(2), B(2), B(1); X5_B(2), X2_B(1), B(1), there exists ring critical points [22] with difference charge density 0.002 and 0.007 $e/Å^3$ respectively (Table 1). For inter-icosahedral interactions [35] like between X3_B(2), B(2), bond critical point exists with difference in charge density value 0.006 $e/Å^3$ (Table 1). However for ring critical points lying between X5_B(2), X3_B(2), B(2); X3_B(2), X4_B(2), X5_B(2), B(2); the difference exists as 0.003 $e/Å^3$. Analysis on *(X4_B(1)-X6-B(1)-X3_B(1))* plane reveals almost no difference in charge density except for the ring critical point formed by X6_B(1), X4_B(1), X3_B(1) atoms with value 0.003 $e$/Å$^3$(Table 1). Analyses of *(X4_B(1)-X6-B(1)-X3_B(1))* plane revealed that while for the ring critical point between atoms X6_B(1)-X4_B(1)-X3_B(1), there exists a difference in charge density value by 0.003 $e$/Å$^3$, that for the bond X6_B(1)-X4_B(1) is zero (Table 1). Figure 3 (e-h) demonstrates the charge density and deformation density maps of specific planes in the α-boron system drawn using X3_B(1), B(1) and X2_B(1) atoms with difference in charge density marked as red contour lines at the surface. Similar charge density maps for plane drawn using X4_B(1), X6-B(1) and X3_B(1) atoms are demonstrated in Figure 4 (a-d).

Similar studies on CaSi were conducted on a plane defined by three atoms [37], namely Ca, X9_Si and X11_Si. For the X9_Si-X11_Si bond, the bond critical point shows a difference in charge density of magnitude 0.005 $e$/Å$^3$ (Table 1). While for the Ca-X11_Si bond the difference stands out to be 0.006 $e$/Å$^3$; for Ca-X9_Si, the difference is 0.008 $e$/Å$^3$ and for Si-X2_Si bond, the difference is 0.010$e$/Å$^3$ (Table 1). Ring critical point for Ca, X9_Si and X11_Si atoms records a charge density difference of 0.003 $e$/Å$^3$ (Table 1). However for the Ca-Si bond, the difference stands out to be zero (upto 3 decimal places). Similar calculations were performed for α-glycine molecule [33] as well. But, for all critical points, the difference in charge density was zero (upto 3 decimal places) (Table 1). However, slight differences between surface and bulk can be seen in the charge density and deformation density maps (Fig. 4). All the above calculations for α-glycine, α-boron, boron carbide and calcium mono-silicide (CaSi) were repeated in IAM mode [38] but in all cases, expectedly no change in charge density values were observed (Table 2). This indicates that for molecular materials, surface truncation effects are least. However, significant differences might be expected for molecular solids where extended network is formed with the aid of various intermolecular interactions, such as hydrogen bonds. The charge density maps highlighting differences are provided in Fig. 4 (e-h) and the Supplementary Information.

The above findings find their interpretation as sources of charge density on a plane when a crystal in continuum encounters a truncation and therefore forms an interface between the bulk crystal and external environment or vacuum that we define as surface [41-42]. As evident from the values of difference in charge density on a plane, the surface in particular has its own charge density picture that involves variable charge densities at different regions of the surface. Since like in bulk, the surface consists of features like bonds, rings, etc. [38], the difference in charge density at the critical points in all these topological features actually contribute in defining the properties of that particular surface [30, 38]. Like for the *(C(4)-X1_B(3)-X5_C(4))* plane in boron carbide, charge densities calculated at bond critical points for C(4)-B(3) bond and X5_C(4)-B(3) bond for contribution from atoms below the plane is smaller than what is calculated for contribution from atoms both above and below the plane (Table 1, Fig. 3). The value is 1.553 $e$/Å$^3$ for C(4)-B(3) bond for atoms below and it is 1.555$e$/Å$^3$ for atoms both above and below the plane. For X5_C(4)-B(3) bond, the values are 1.522 $e$/Å$^3$ and 1.555 $e$/Å$^3$ respectively. This suggests that the surface defined by above plane bears a deficiency of electrons (Table 1, Fig. 3).

In order to check the practical applicability of surface charge density using our method, we have performed chemiresistive gas sensing with boron carbide at high temperatures (see Supplementary Information). The sample showed a steady p-type response to ammonia (Fig. 5). Hence when a reducing gas like ammonia interacts with the chemisorbed oxygen layer on boron carbide surface, there is a transfer of electrons back into the boron carbide system [14, 43]. As can be seen from figure 3, the surface containing *(C(4)-X1_B(3)-X5-C(4))* plane is substantially more electron deficient than the bulk. Thus the ammonia molecules shall have greater chance of donating electrons to the system via mentioned surface. This leads to a rise in electrical resistance of the typical p-type boron carbide system, making the material sense reducing gas like ammonia (Fig. 5) [14, 43]. Moreover it is evident that while some surfaces under consideration posses these differences in charge densities, several others don't posses that (Table 1, Fig. 4 and Supplementary Information). This can be anticipated to be one of the reasons why certain surfaces in materials are found to be experimentally active for certain external phenomenon while some are not (Fig. 5) [44]. This also bears the cause behind differential behavior of different material planes under external factors. Observance of these differences on considering the multipole atom model categorically substantiates the necessity of considering atom-atom interactions in order to define an entity on some surface (Table 1 and 2).

The above discussion explains why surface behaves differently than the bulk, as evident from numerous surface dependent phenomena like gas sensing, catalysis, etc. [1-10]. Since the contribution of surrounding atoms depend on the crystal structures, chemical interaction/bonding and atomic planes, it is obvious that the difference between bulk and surface will be different depending on which plane has been truncated (Table 1). For compounds with heavy element, modelling the outer core electrons besides valence electrons will provide more accurate picture of charge density distribution [45]. This mechanism of elucidating surface structure from bulk crystal structure will not only open the possibility to determine surface structure from bulk crystal structure but also enable researchers to study surface using the popular technique of XRD. This technique avoids involvement of sophisticated experimental tools specialized for surface analyses and ensures charge density estimation accurately upto two decimal places in $e/Å^3$, yet unachieved by methods adopted till date. The generality of this method ensures that this method will be applicable to theoretically calculated charge densities or invariom model densities [46]. As the surface in this work has been defined as ideally truncated plane with respect to vacuum (Fig. 1), the surface charge density picture may vary based on post truncation internal and external interactions. Including charge density contributions of the atoms of surface layers towards the atoms that were above those layers before truncation from the bulk, back to the surface layer atoms by source function calculation and using other complementary methods could yield more accurate picture of surface structure [25-26]. Further developments using dynamic charge density [33, 47] and source function calculations [26] for extended solids shall provide more information and enable evolution of greater insights into surface structure determination from bulk X-ray diffraction data.

## Conclusion

In this work, using the concept of aspherical charge density models of bulk crystal structures in different molecular and extended solids, we have shown a convenient way of determining high-resolution surface structures both in qualitative and quantitative manner from bulk X-ray diffraction data. Differences in charge density at different critical points on the planes in different molecular and extended solids were calculated, based on presence and absence of atoms above the plane. While for molecular solids like α-glycine no such detectable difference was observed, for extended solids like α-boron, boron carbide and CaSi, significant differences in charge density distributions between bulk and surface were found. This was observed only when atom-atom interactions were considered while calculating charge densities at different parts of the plane, highlighting importance of considering inter-atomic interactions for defining a surface. While for some planes this difference is significant, it is negligible for others. This indicates a plausible reason for different behavior of one surface in a material than the other. The magnitude and nature of charge densities calculated are expected to delineate the surface structure which can be anticipated to explain significant surface phenomena.

## Acknowledgements

Authors acknowledge Prof. Wolfgang Scherer, Universität Augsburg, Germany for the bulk charge density model of CaSi and the Centre for Research in Nanoscience and Nanotechnology, University of Calcutta for TEM facility. The authors thank Sander van Smaalen, University of Bayreuth, Germany for discussions. N. Chakraborty would like to acknowledge DST, Govt. of India for the INSPIRE fellowship (IF170810). S. Mondal would like to acknowledge financial support from SERB Core Research Grant, Government of India (Grant number: CRG/2019/004588).

**Author Contributions** S. M. conceived and supervised the research project. N. C. carried out surface charge density determination with the assistance of S. M. Gas sensing experiments were carried out by N. C. S. M. is responsible for the bulk charge density models of α-glycine, α-boron and boron carbide. N. C. and S. M. analyzed the experimental results and wrote the manuscript.

**Competing interests** The authors declare no competing interests.

## References

[1] Boukhvalov, D. W., Paolucci, V., D'Olimpio, G., Cantalinic, C. & Politano, A. Chemical reactions on surfaces for applications in catalysis, gas sensing, adsorption-assisted desalination and Li-ion batteries: opportunities and challenges for surface science. Phys. Chem. Chem. Phys. 23, 7541-7552 (2021).

[2] Chae, H., Siberio-Pérez, D., Kim, J., Go, Y., Eddaoudi, M., Matzger, A. J., O'Keeffe, M., Yaghi, O. M. & Materials Design and Discovery Group A route to high surface area, porosity and inclusion of large molecules in crystals. Nature 427, 523–527 (2004).

[3] Oosthuizen, D. N., Motaung, D. E. & Swart, H. C. Gas sensors based on $CeO_2$ nanoparticles prepared by chemical precipitation method and their temperature-dependent selectivity towards $H_2S$ and $NO_2$ gases. Applied Surface Science 505, 1, 144356 (2020).

[4] Chakraborty, N. & Mondal, S. Dopant Mediated Surface Charge Imbalance in Enhancing Performance of Metal Oxide Chemiresistive Gas Sensors. J. Mater. Chem. C 10, 1968-1976 (2022).

[5] Batzill, M. & Diebold, U. Surface studies of gas sensing metal oxides. Phys. Chem. Chem. Phys. 9, 2307-2318 (2007).

[6] Sauer, T., Cesconeto Neto, G., José, H. J. & Moreira, R. F. P. M. Kinetics of photocatalytic degradation of reactive dyes in a $TiO_2$ slurry reactor. Journal of Photochemistry and Photobiology A: Chemistry 149(1-3), 147-154 (2002).

[7] Song, I., Lee, H., Jeon, S. W., Ibrahim, I. A. M., Kim, J., Byun, Y., Koh, D. J., Han, J. W. & Kim, D. H. Simple physical mixing of zeolite prevents sulfur deactivation of vanadia catalysts for NOx removal. Nat. Commun. 12, 901 (2021).

[8] Daniel, M. C. &Astruc, D. Gold Nanoparticles: Assembly, Supramolecular Chemistry, Quantum-Size-Related Properties, and Applications toward Biology, Catalysis, and Nanotechnology. Chem. Rev. 104, 293−346 (2004).

[9] Leblebici, S. Y., Leppert, L., Li, Y., Reyes-Lillo, S. E., Wickenburg, S., Wong, E., Lee, J., Melli, M., Ziegler, D., Angell, D. K., Frank Ogletree, D., Ashby, P. D., Toma, F. M., Neaton, J. B., Sharp, I. D. & Weber-Bargioni, A. Facet-dependent photovoltaic efficiency variations in single grains of hybrid halide perovskite. Nat Energy 1, 16093 (2016).

[10] Lei, W., Portehault, D., Liu, D., Qin, S. & Chen, Y. Porous boron nitride nanosheets for effective water cleaning. Nat. Commun. 4, 1777 (2013).

[11] Lüth, H. (2001) Surface and Interface Physics: Its Definition and Importance. In: Solid Surfaces, Interfaces and Thin Films. Advanced Texts in Physics. Springer, Berlin, Heidelberg

[12] Surface Characterization: A User's Sourcebook Edited by Brune, D., Hellborg, R., Whitlow, H. J. & Hunderi, O. WILEY-VCH Verlag.

[13] Woodruff, D. (2016). Methods of Surface Structure Determination. In Modern Techniques of Surface Science (pp. 98-214). Cambridge: Cambridge University Press. doi:10.1017/CBO9781139149716.006

[14] Chakraborty, N., Sanyal, A., Das, S., Saha, D., Medda, S. K. & Mondal, S. Ammonia Sensing by $Sn_{1-x}V_xO_2$ Mesoporous Nanoparticles. ACS Appl. Nano Mater. 3, 8, 7572-7579 (2020).

[15] Feidenhans'l, R. Surface structure determination by X-ray diffraction, Surface Science Reports, 10, 3, 105-188 (1989).

[16] Harris, K. D. M. & Tremayne, M. Crystal Structure Determination from Powder Diffraction Data. Chem. Mater. 8, 11, 2554–2570 (1996).

[17] Harris, K. D. M., Tremayne, M. & Kariuki, B. M. Contemporary Advances in the Use of Powder X-Ray Diffraction for Structure Determination. Angew. Chem. Int. Ed. 40, 1626-1651 (2001).

[18] Warren, B. E. X-Ray Diffraction (Addison-Wesley, Reading, MA, 1969).

[19] Gustafson, J., Shipilin, M., Zhang, C., Stierle, A., Hejral, U., Ruett, U., Gutowski, O., Carlsson, P. A., Skoglundh, M. & Lundgren, E. High-Energy Surface X-ray Diffraction for Fast Surface Structure Determination. Science 343, 758-761 (2014).

[20] Rous, P. J., Pendry, J. B. & Saldin, D. K. Tensor LEED: A Technique for High-Speed Surface-Structure Determination. PHYSICAL REVIEW LETTERS 57(23), 2951-2954 (1986).

[21] Heinz, K., Saldin, D. K. & Pendry, J. B. Diffuse LEED and Surface Crystallography. PHYSICAL REVIEW LETTERS 55(21), 2312-2315 (1985).

[22] Gao, W., Addiego, C., Wang, H., Yan, X., Hou, Y., Ji, D., Heikes, C., Zhang, Y., Li, L., Huyan, H., Blum, T., Aoki, T., Nie, Y., Schlom, D. G., Wu, R. & Pan, X. Real-space charge-density imaging with sub-ångström resolution by four-dimensional electron microscopy. Nature 575, 480–484 (2019).

[23] Renaud, G., Lazzari, R. & Leroy, F. Probing surface and interface morphology with Grazing Incidence Small Angle X-Ray Scattering. Surface Science Reports 64, 255–380 (2009).

[24] Rauscher, M., Paniago, R., Metzger, H., Kovats, Z., Domke, J., Peisl, J., Pfannes, HD., Schulze, J. and Eisele, I. Grazing incidence small angle x-ray scattering from free-standing nanostructures. J. Appl. Phys. 86, 12, 6763-6769 (1999).

[25] Bader, R. F. W. & Gatti, C. A Green's function for the density. Chemical Physics Letters 287, 233–238 (1998).

[26] Gatti, C., Saleh, G. & Presti, L. L. Source Function applied to experimental densities reveals subtle electron-delocalization effects and appraises their transferability properties in crystals. Acta Cryst. B 72, 180–193 (2016).

[27] Hu, B., McCandless, G. T., Menard, M., Nascimento, V. B., Chan, J. Y., Plummer, E. W. & Jin, R. Surface and bulk structural properties of single-crystalline $Sr_3Ru_2O_7$. PHYSICAL REVIEW B 81, 184104 (2010).

[28] Cullity, B. D. & Stock, S. R. Elements of X-ray Diffraction, Addison-Wesley Publishing Company, Inc. (1956).

[29] Dittrich, B., Weber, M., Kalinowski, R., Grabowsky, S., Hubschlea, C. B. & Luger, P. How to easily replace the independent atom model– the example of bergenin, a potential anti-HIV agent of traditional Asian medicine. Acta Cryst. B 65, 749–756 (2009).

[30] Coppens, P. (1997). X-ray Charge Densities and Chemical Bonding. New York: Oxford University Press.

[31] Dittrich, B., Hubschle, C. B., Messerschmidt, M., Kalinowski, R., Girnt, D. and Luger, P. The invariom model and its application: refinement of D,L-serine at different temperatures and resolution. Acta Cryst. A 61, 314–320 (2005).

[32] Stewart, R. F. & Spackman, M. A. (1983). VALRAY Users Manual. Department of Chemistry, Carnegie-Mellon University, Pittsburgh, USA.

[33] Mondal, S., Prathapa, S. J. & van Smaalen, S. Experimental dynamic electron densities of multipole models at different temperatures. Acta Cryst. A 68, 568-581 (2012).

[34] Destro, R., Roversi, P., Barzaghi, M.& Marsh, R. E. Experimental Charge Density of α-Glycine at 23 K. J. Phys. Chem. A, 104, 1047–1054 (2000).

[35] Mondal, S., van Smaalen, S., Parakhonskiy, G., Jagannatha Prathapa, S., Noohinejad, L., Bykova, E. &Dubrovinskaia, N. Experimental evidence of orbital order in α-B12 and γ-B28 polymorphs of elemental boron. PHYSICAL REVIEW B 88, 024118 (2013).

[36] Mondal, S., Bykova, E., Dey, S., Imran Ali, S., Dubrovinskaia, N., Dubrovinsky, L., Parakhonskiy, G. & van Smaalen, S. Disorder and defects are not intrinsic to boron carbide. Sci. Rep. 19330 (2016).

[37] Kurylyshyn, I. M., Fassler, T. F., Fischer, A., Hauf, C., Eickerling, G., Presnitz, M.& Scherer, W. Probing the Zintl–Klemm Concept: A Combined Experimental and Theoretical Charge Density Study of the Zintl Phase CaSi. Angew. Chem. Int. Ed. 53, 3029 –3032 (2014).

[38] Bader, R. F. W. (1990). Atoms in Molecules – a Quantum Theory. New York: Oxford University Press.

[39] Volkov, A., Macchi, P., Farrugia, L. J., Gatti, C., Mallinson, P., Richter, T. & Koritsanszky, T. (2006). XD2006, A Computer Program Package for Multipole Refinement, Topological Analysis of Charge Densities and Evaluation of Intermolecular Energies from Experimental or Theoretical Structure Factors. URL http://xd.chem.buffalo.edu/

[40] Mondal, S. Charge Transfer and Fractional Bonds in Stoichiometric Boron Carbide. Chem. Mater. 29, 6191−6194 (2017).

[41] Tran, R., Xu, Z., Radhakrishnan, B., Winston, D., Sun, W., Persson, K. A. & Ong, S. P. Data Descriptor: Surface energies of elemental crystals. Sci Data 3, 160080 (2016).

[42] Bare, S. R. & Somorjai, G. A. Surface Chemistry, Editor(s): Robert A. Meyers, Encyclopaedia of Physical Science and Technology (Third Edition), Academic Press 373-421 (2003).

[43] Ding, Y., Guo, X., Du, B., Hu, X., Yang, X., He, Y., Zhou, Y. & Zang, Z. Low-operating temperature ammonia sensor based on $Cu_2O$ nanoparticles decorated with p-type $MoS_2$ nanosheets. J. Mater. Chem. C 9, 4838 (2021).

[44] Kaneti, Y. V., Zhang, Z., Yue, J., Zakaria, Q. M. D., Chen, C., Jiang, X. & Yu, A. Crystal plane-dependent gas-sensing properties of zinc oxide nanostructures: experimental and theoretical studies. Phys. Chem. Chem. Phys. 16, 11471 (2014).

[45] Mondal, S. Experimental Charge density Studies of Inorganic Solids in: Understanding Intermolecular Interactions in the Solid State: Approaches and Techniques. Monographs in Supramolecular Chemistry. Royal Society of Chemistry 130−158 (2018).

[46] Dittrich, B., Koritsanszky, T. & Luger, P. A Simple Approach to Nonspherical Electron Densities by Using Invarioms. Angew. Chem. Int. Ed. 43, 2718–2721 (2004).

[47] Prathapa, S. J., Mondal, S. & van Smaalen S. Electron densities by the maximum entropy method (MEM) for various types of prior densities: a case study on three amino acids and a tripeptide. Acta Cryst. B 69, 203–213 (2013).

# Tables and figures

Table 1: Comparison of charge density values at different bond critical points for α-glycine, α boron, boron carbide and CaSi using multipole (MP) formalism.

| bonds/rings | $\rho_{base}(e/\text{Å}^3)$ | $\rho_{base+top}(e/\text{Å}^3)$ | $\Delta\rho(e/\text{Å}^3)$ |
|---|---|---|---|
| *α boron (X3_B(1)-B(1)-X2_B(1))* | | | |
| Intra-icosahedral: | | | |
| X2_B(1)-X3_B(1)-B(1) | 0.776 | 0.790 | 0.014 |
| X3_B(1)-B(1) | 0.800 | 0.816 | 0.016 |
| X2_B(1)-B(1) | 0.800 | 0.816 | 0.016 |
| X2_B(1)-X3_B(1) | 0.800 | 0.816 | 0.016 |
| B(1)-B(2) | 0.756 | 0.760 | 0.004 |
| X5_B(2)-B(2) | 0.802 | 0.797 | 0.005 |
| X5_B(2)-B(2)-B(1) | 0.709 | 0.711 | 0.002 |
| X5_B(2)-X2_B(1)-B(1) | 0.693 | 0.700 | 0.007 |
| | | | |
| Inter-icosahedral: | | | |
| X3_B(2)-B(2) | 0.559 | 0.553 | 0.006 |
| X5_B(2)-X3_B(2)-B(2) | 0.237 | 0.240 | 0.003 |
| X3_B(2)-X4_B(2)-X5_B(2)-B(2) | 0.237 | 0.240 | 0.003 |
| *α boron (X4_B(1)-X6-B(1)-X3_B(1))* | | | |
| X6_B(1)-X4_B(1)-X3_B(1) | 0.104 | 0.107 | 0.003 |
| X6_B(1)-X4_B(1) | 0.816 | 0.816 | 0.000 |
| *boron carbide (C(4)-X1_B(3)-X5-C(4))* | | | |
| C(4)-B(3) | 1.553 | 1.555 | 0.002 |
| X5_C(4)-B(3) | 1.522 | 1.555 | 0.033 |
| X1_C(4)-X1_B(3) | 1.553 | 1.555 | 0.002 |
| X1_C(4)-B(3) | 0.122 | 0.124 | 0.002 |
| *boron carbide (X4_B(2)-X5_B(2)-X6_B(2))* | | | |
| X4_B(2)-X5_B(2) | 0.722 | 0.722 | 0.000 |
| X4_B(2)-X5_B(2)-X6_B(2) | 0.631 | 0.632 | 0.001 |
| X5_B(2)-X6_B(2) | 0.722 | 0.722 | 0.000 |
| *CaSi (X9_Si(1)-Ca(1)-X11_Si(1))* | | | |
| X9_Si-X11_Si | 0.455 | 0.460 | 0.005 |
| Ca-X11_Si | 0.106 | 0.112 | 0.006 |
| Ca-X9_Si | 0.115 | 0.123 | 0.008 |
| Ca-X9_Si-X11_Si | 0.109 | 0.112 | 0.003 |
| Si-X2_Si | 0.450 | 0.460 | 0.010 |
| Ca-Si | 0.143 | 0.143 | 0.000 |
| *α-glycine (C(1)-C(2)-O(2))* | | | |
| C(1)-C(2) | 1.735 | 1.735 | 0.000 |
| C(1)-O(1) | 2.770 | 2.770 | 0.000 |
| C(1)-O(2) | 2.733 | 2.733 | 0.000 |
| C(2)-N | 1.691 | 1.691 | 0.000 |

*Symmetry related atomic co-ordinates for B sites in α boron: X2 –y, x-y, z; X3 -x+y,-x,z; X4 y,x,-z; X5 x-y,-y,-z; X6 -x,-x+y,-z*

*Symmetry related atomic co-ordinates for B sites in boron carbide: X1 x,y,z; X4 y,x,-z; X5 x-y,-y,-z; X6 –x,-x+y,-z*

*Symmetry related atomic co-ordinates for Si sites in CaSi: X2 –x,-y,1/2+z; X9 1/2+x,1/2+y,z; X11 1/2+x,1/2-y,-z*

Table 2: Comparison of charge density values at different bond critical points for α-glycine, α boron, boron carbide and CaSi using independent atom model (IAM).

| bonds/rings | $\rho_{base}(e/Å^3)$ | $\rho_{base+top}(e/Å^3)$ | $\Delta\rho(e/Å^3)$ |
|---|---|---|---|
| *α boron (X3_B(1)-B(1)-X2_B(1))* | | | |
| Intra-icosahedral: | | | |
| X2_B(1)-X3_B(1)-B(1) | 0.628 | 0.628 | 0.000 |
| X3_B(1)-B(1) | 0.673 | 0.673 | 0.000 |
| X2_B(1)-B(1) | 0.673 | 0.673 | 0.000 |
| X2_B(1)-X3_B(1) | 0.673 | 0.673 | 0.000 |
| B(1)-B(2) | 0.636 | 0.636 | 0.000 |
| X5_B(2)-B(2) | 0.645 | 0.645 | 0.000 |
| X5_B(2)-B(2)-B(1) | 0.305 | 0.305 | 0.000 |
| Inter-icosahedral: | | | |
| X3_B(2)-B(2) | 0.448 | 0.448 | 0.000 |
| *α boron (X4_B(1)-X6-B(1)-X3_B(1))* | | | |
| X6_B(1)-X4_B(1) | 0.673 | 0.673 | 0.000 |
| *boron carbide(C(4)-X1_B(3)-X5-C(4))* | | | |
| C(4)-B(3) | 0.958 | 0.958 | 0.000 |
| X5_C(4)-B(3) | 0.943 | 0.943 | 0.000 |
| X1_C(4)-X1_B(3) | 0.958 | 0.958 | 0.000 |
| X1_C(4)-B(3) | 0.063 | 0.063 | 0.000 |
| *boron carbide (X4_B(2)-X5_B(2)-X6_B(2))* | | | |
| X4_B(2)-X5_B(2) | 0.621 | 0.621 | 0.000 |
| X4_B(2)-X5_B(2)-X6_B(2) | 0.564 | 0.564 | 0.000 |
| X5_B(2)-X6_B(2) | 0.621 | 0.621 | 0.000 |
| *α-glycine (C(1)-C(2)-O(2))* | | | |
| C(1)-C(2) | 1.190 | 1.190 | 0.000 |
| C(1)-O(1) | 2.041 | 2.041 | 0.000 |
| C(1)-O(2) | 2.034 | 2.034 | 0.000 |
| C(2)-N | 1.405 | 1.405 | 0.000 |

*Symmetry related atomic co-ordinates for B sites in α boron: X2 –y, x-y, z; X3 -x+y,-x,z; X4 y,x,-z; X5 x-y,-y,-z; X6 -x,-x+y,-z*

*Symmetry related atomic co-ordinates for B sites in boron carbide: X1 x,y,z; X4 y,x,-z; X5 x-y,-y,-z; X6 –x,-x+y,-z*

*Symmetry related atomic co-ordinates for Si sites in CaSi: X2 –x,-y,1/2+z; X9 1/2+x,1/2+y,z; X11 1/2+x,1/2-y,-z*

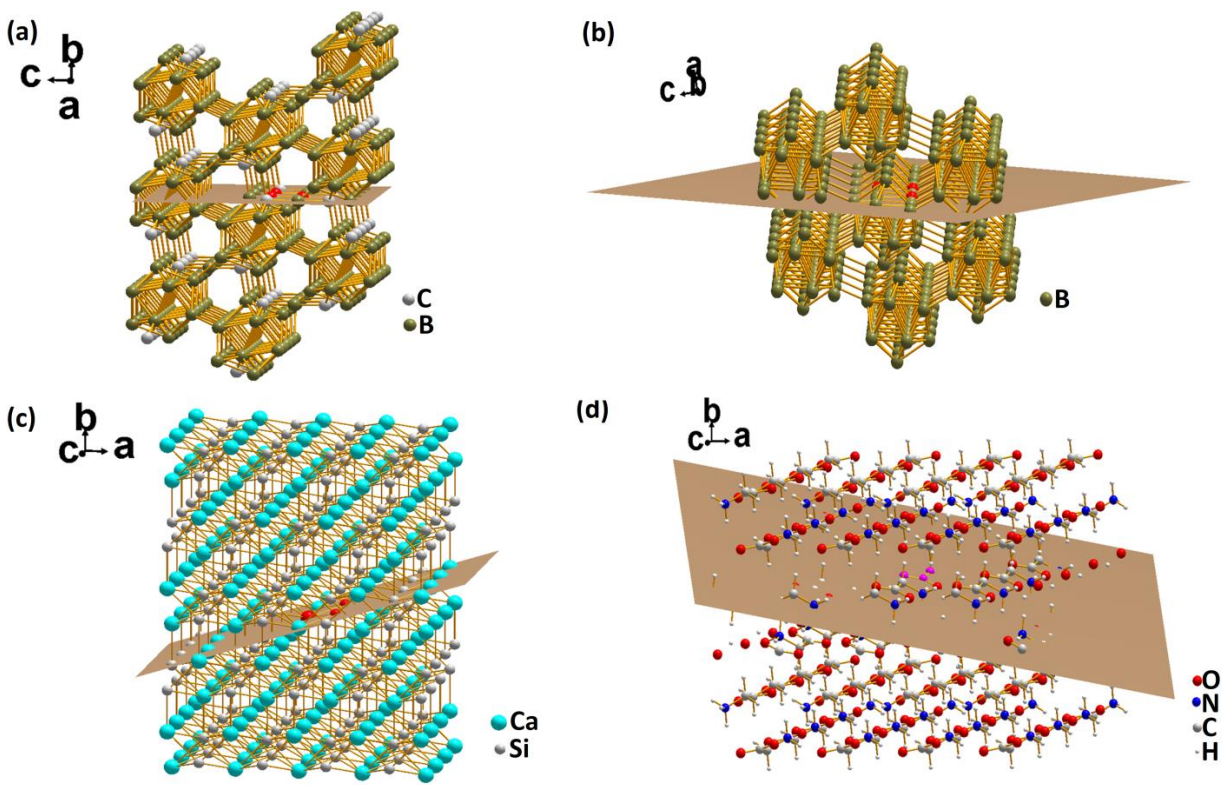

**Fig. 1:** Scheme for drawing planes with atoms above and below the plane (a) boron carbide [(C(4)-X1_B(3)-X5-C(4)) corresponding to (0 1 0) plane] (b) α boron [(X3_B(1)-B(1)-X2_B(1)) corresponding to (1 0 0) plane] (c) CaSi (X9_Si(1)-Ca(1)-X11_Si(1)) corresponding to (-1 2 0) plane] and (d) α-glycine [(C(1)-C(2)-O(2)) corresponding to (1 -1 3) plane].

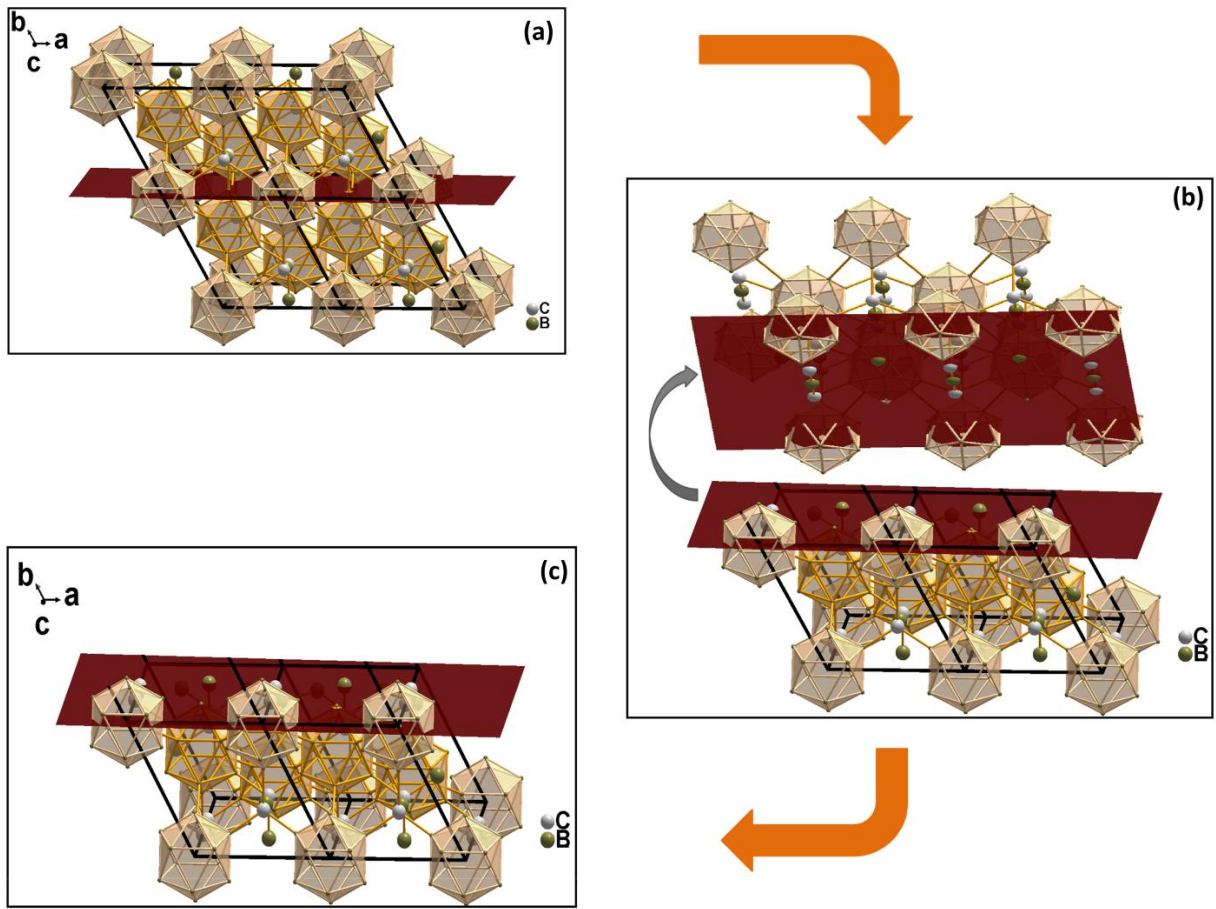

**Fig.2:** Schematic diagram delineating methodology of defining surface. (a) A cut introduced through the bulk. (b) Removal of the region above the cut. (c) The plane (0 1 0) exposed is now defined as surface.

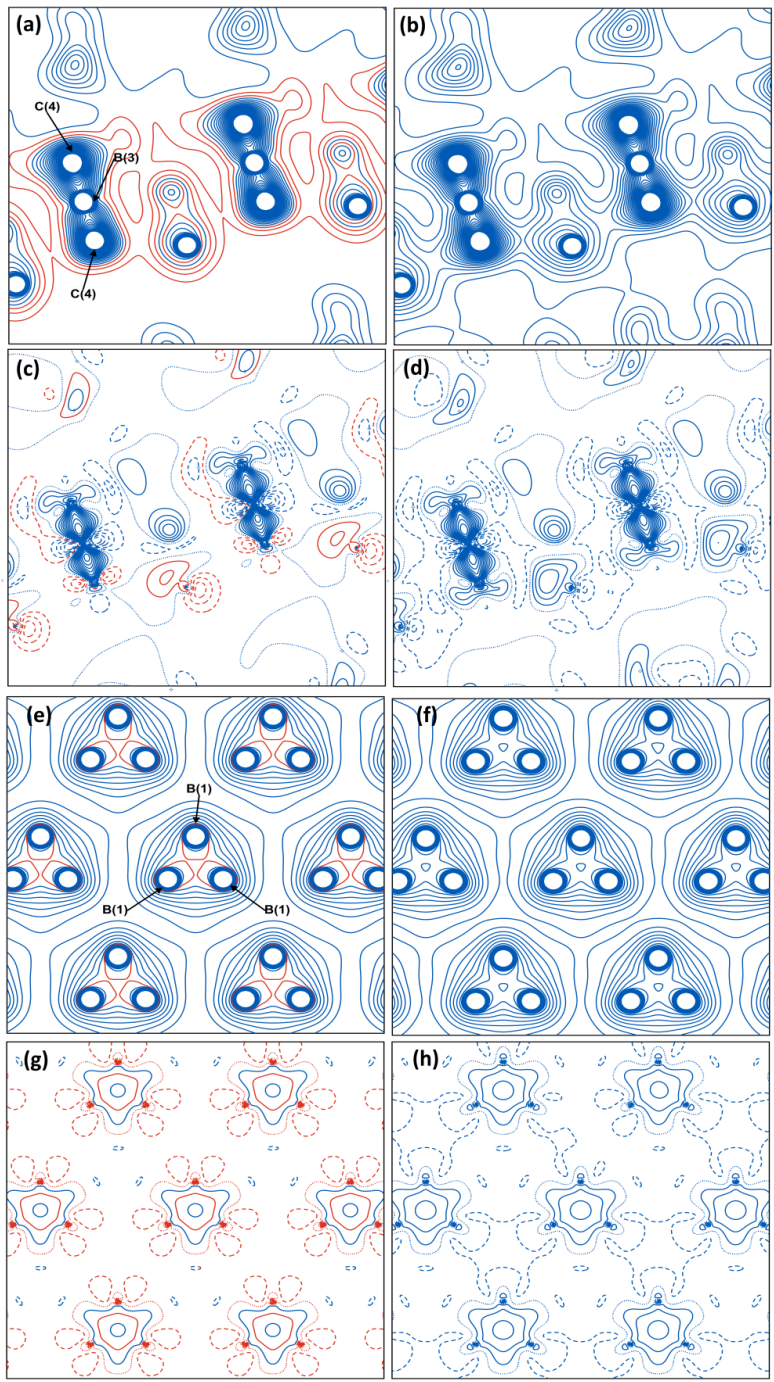

**Fig. 3:** Comparison of (a-b, e-f) charge density (contours at 0.1upto2.5 electron Å$^{-3}$) and (c-d, g-h) deformation density maps (contours at 0.05 upto 1.0 electron Å$^{-3}$) in boron carbide *(C(4)-X1_B(3)-X5-C(4))* [(01 0) plane] and α-boron *(X3_B(1)-B(1)-X2_B(1))* [(0 0 -1) plane] respectively. (a, c) Plane containing B(1), B(3) and C(4) atoms representing surface (b, d) bulk. Plane containing B(1) atoms representing (e, g) surface (f, h) bulk. The contour lines on surface different from bulk are separately shown in red color. Solid contour lines are for positive values, dashed lines for negative values, and dotted lines for zero contour.

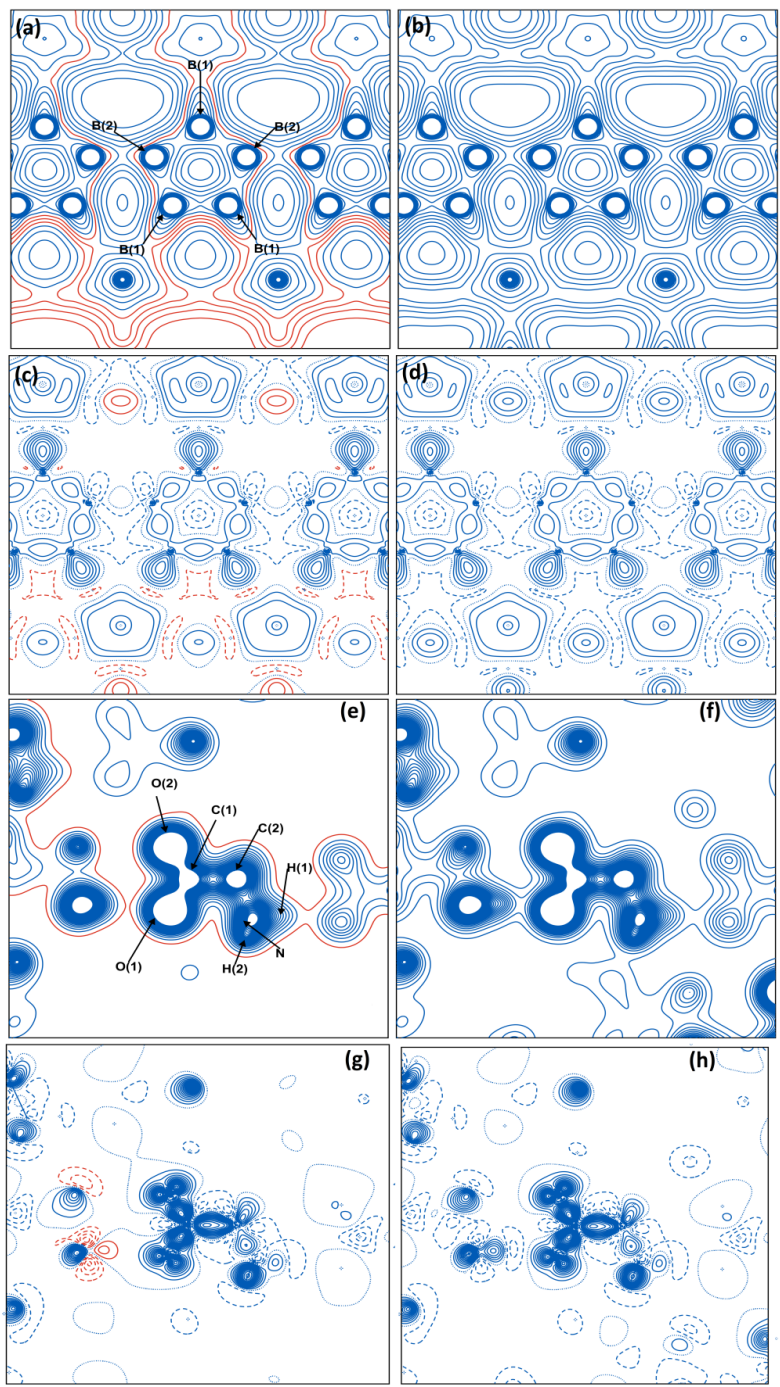

**Fig. 4:** Comparison of (a-b, e-f) charge density (contours at 0.1 upto 2.5 electron Å$^{-3}$) and (c-d, g-h) deformation density maps (contours at 0.05 upto 1.0 electron Å$^{-3}$) in α-boron *(X4_B(1)-X6-B(1)-X3_B(1))*[(1 0 0) plane] and α-glycine *(C(1)-C(2)-O(2))*[(1 -1 3) plane] respectively. (a, c) Plane containing B(1), B(2) atoms representing surface (b, d) bulk. Plane containing C(1), C(2), O(2) atoms representing (e, g) surface (f, h) bulk. The contour lines on surface different from bulk are separately shown in red color. Solid contour lines are for positive values, dashed lines for negative values, and dotted lines for zero contour.

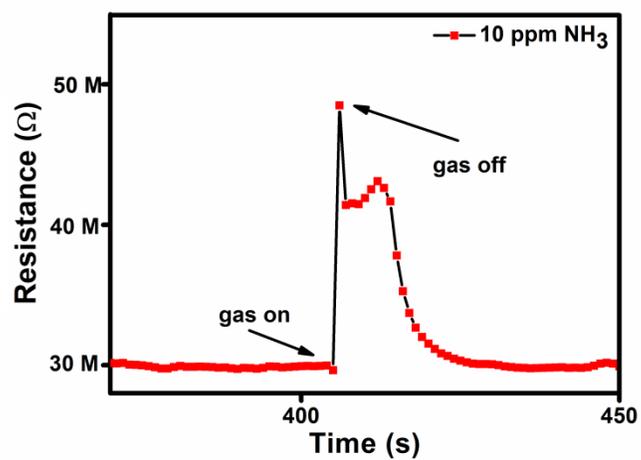

**Fig. 5:** Dynamic sensing characteristics of boron carbide sensor to 10 ppm NH$_3$ at an elevated temperature of 600°C.

# Supplementary Information

# High-resolution surface structure determination from bulk X-ray diffraction data

Nirman Chakraborty and Swastik Mondal[*]

*CSIR-Central Glass and Ceramic Research Institute, 196, Raja S. C. Mullick Road, Jadavpur, Kolkata-700032, India*

[*]Corresponding author: swastik_mondal@cgcri.res.in

**Gas sensing experiments**

For developing the gas sensor, 0.1 g of boron carbide powder was thoroughly mixed in iso-propyl alcohol into a consistent slurry and drop-casted on a flat alumina substrate with interdigitated electrodes made using screen printing technique. The coating was allowed to dry at 60°C for 6 hours and then connections were made using platinum wires. The whole sensor along with the wires was placed in a tube furnace and the sensor resistance was measured by a $7\tfrac{1}{2}$ digit multimeter (Keysight 34470A) interfaced with the Keysight GUI software. For sensing purposes, standard calibrated cylinders of ammonia in various ppm concentrations were used. They were connected using a Mass Flow Controller (MFC) with gas flow regulated at 50 sccm/min. Sensing experiments were conducted at different temperatures with $NH_3$ gas flow by operating the tube furnace. Optimum sensing response was recorded at 600°C for 10 ppm ammonia gas in air medium.

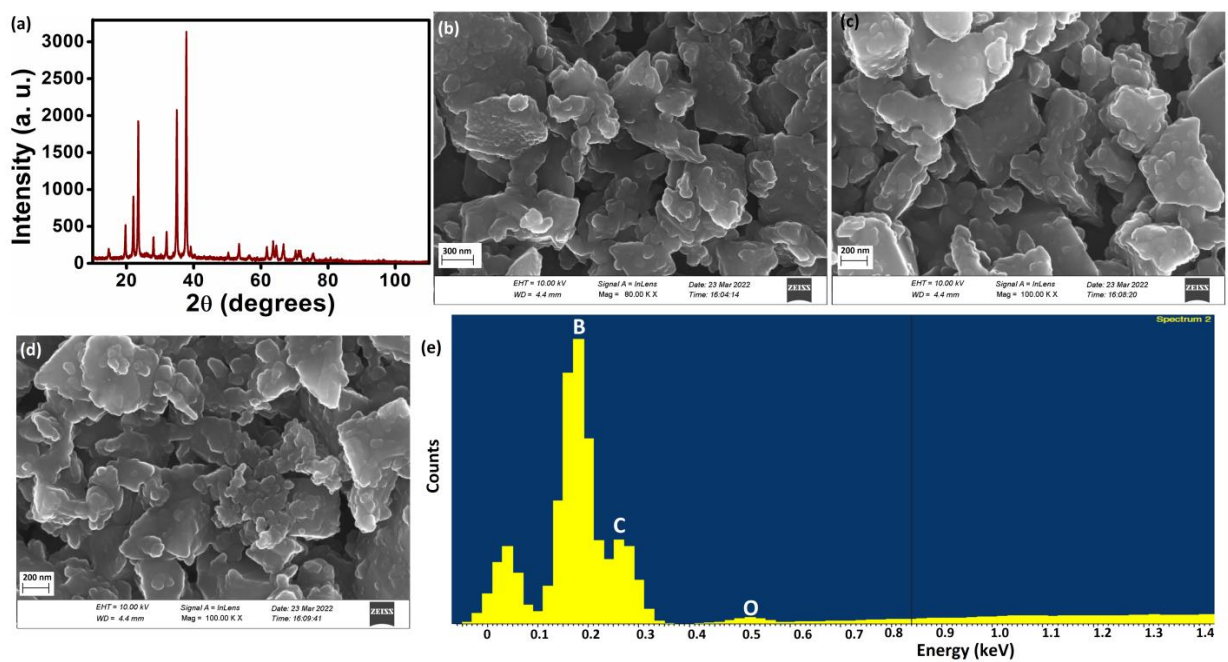

Fig. S1: (a) Powder XRD pattern (b-d) Surface morphology and (e) Surface elemental composition using Energy Dispersive X-ray (EDX) spectroscopy of boron carbide powder used for gas sensing application.

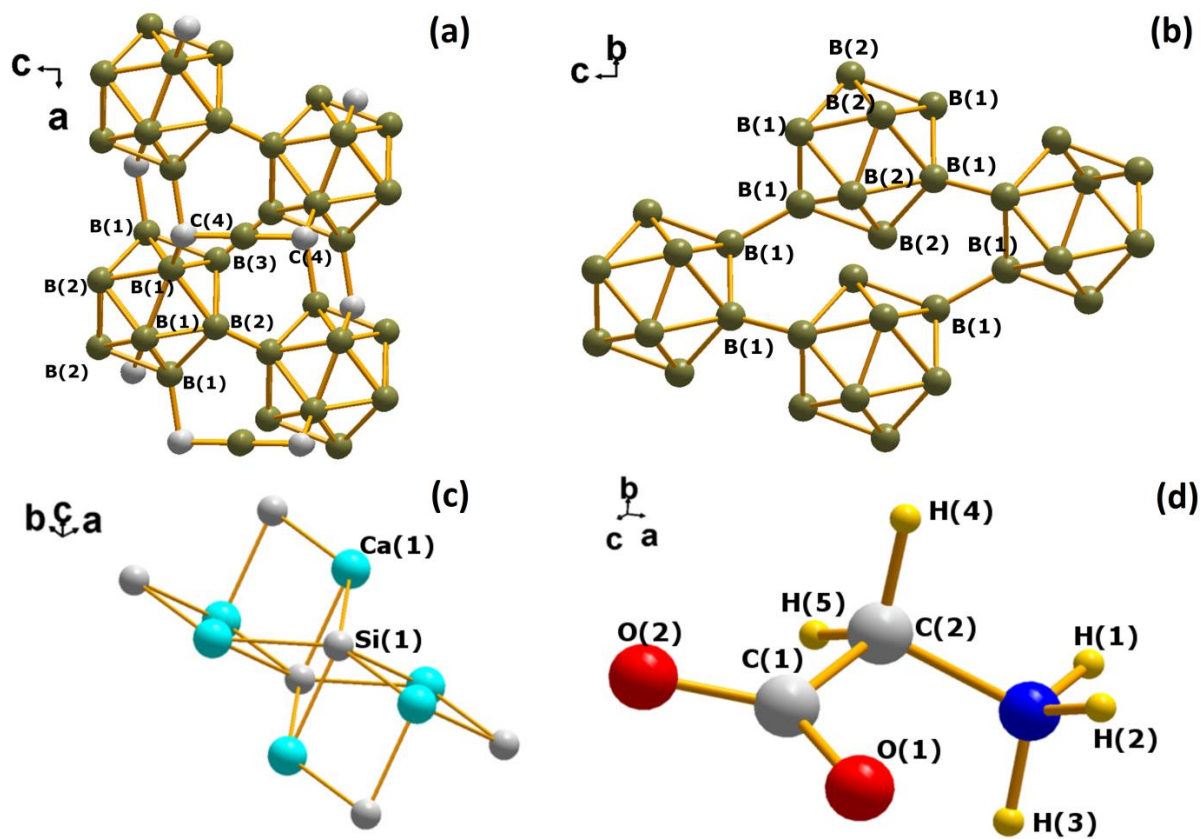

Fig. S2: Crystal structures of (a) boron carbide (b) α-boron (c) calcium monosilicide and (d) α-glycine respectively, identifying the different atoms in crystal.

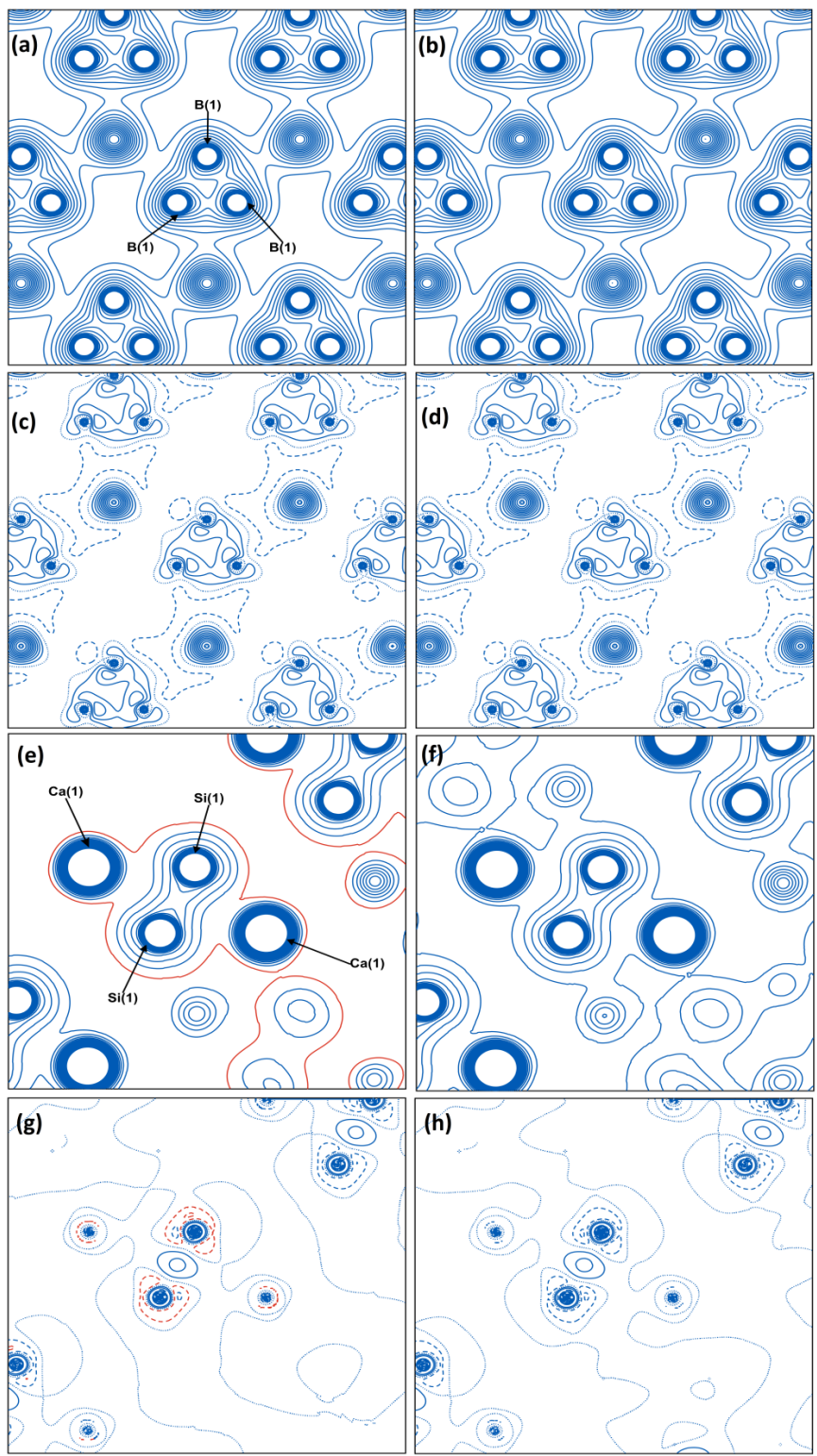

Fig. S3: Comparison of (a-b, e-f) electron density (contours at 0.1 upto 2.5 electron Å$^{-3}$) and (c-d, g-h) deformation density maps (contours at 0.05 upto 1.0 electron Å$^{-3}$) in boron carbide *(X4_B(2)-X5_B(2)-X6_B(2))* [(-1 1 4) plane] and CaSi *(X9_Si(1)-Ca(1)-X11_Si(1))* [(-1 2 0)

plane] respectively. (a, c) Plane containing B(2) atoms representing surface (b, d) bulk. Plane containing Si(1), Ca(1) atoms representing (e, g) surface (f, h) bulk. The contour lines on surface different from bulk are separately shown in red color. Solid contour lines are for positive values, dashed lines for negative values, and dotted lines for zero contours.

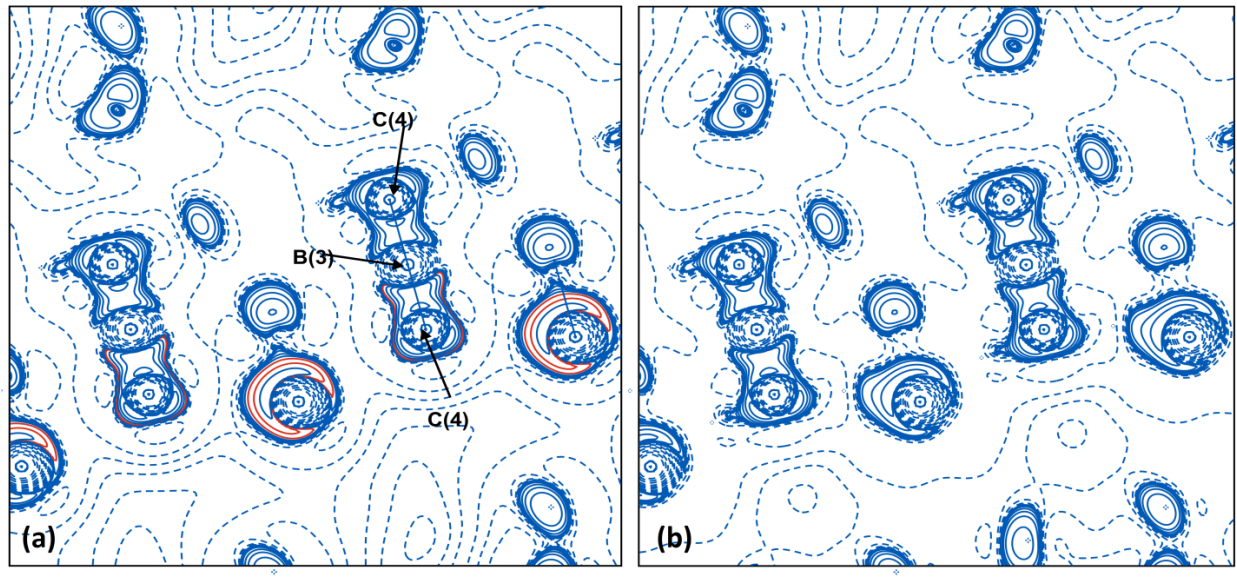

Fig. S4: Comparison of (a-b) Laplacian of charge density maps [contours at $\pm(2, 4, 8) \times 10^n$ electron Å$^{-5}$ ($-3 \leq n \leq 3$)] in boron carbide *(C(4)-X1_B(3)-X5-C(4))*. (a) Plane containing B(1), B(3) and C(4) atoms representing surface (b) bulk. The contour lines on surface different from bulk are separately shown in red color. Solid contour lines are for positive values, dashed lines for negative values.

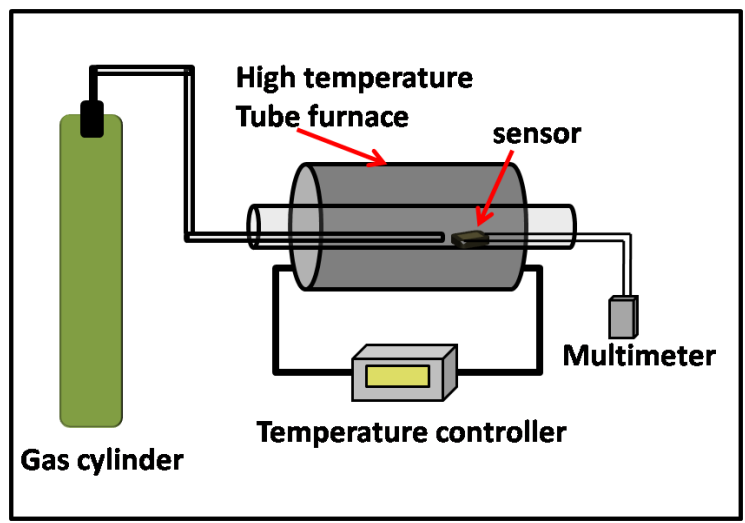

Fig. S5: High temperature ammonia sensing set up for studying surface properties of boron carbide in presence of ammonia gas.

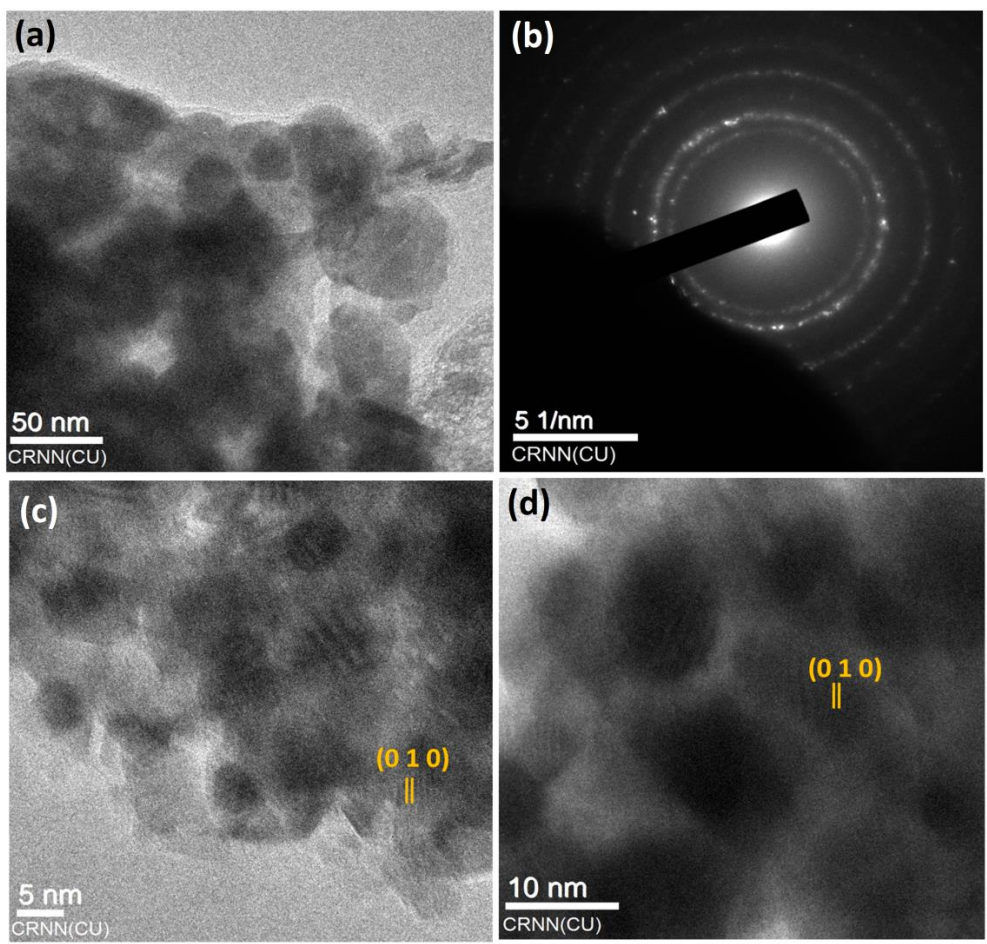

Fig. S6: (a) Bright field Transmission Electron Microscopy (TEM) (b) Selected Area Electron Diffraction (SAED) pattern (c, d) High resolution TEM images of boron carbide powder used for sensing application.